\documentclass[aps,pra,twocolumn,groupedaddress]{revtex4-1}

\usepackage{graphicx}
\usepackage{amsmath} % AMS Math Package
\usepackage{amssymb}	% Math symbols such as \mathbb
\usepackage{epstopdf}
\usepackage{hyperref}
\hypersetup{
    colorlinks=true,       % false: boxed links; true: colored links
    linkcolor=black,          % color of internal links (change box color with linkbordercolor)
    citecolor=black,        % color of links to bibliography
    filecolor=magenta,      % color of file links
    urlcolor=blue           % color of external links
}

\begin{document}
	
	\title{Nonadiabatic dynamics and multiphoton resonances in strong field molecular ionization with few cycle laser pulses}
	
	\author{Vincent Tagliamonti,$^{1}$ P\'eter S\'andor,$^{1}$ Arthur Zhao,$^{1}$ Tam\'as Rozgonyi,$^{2}$ Philipp Marquetand,$^{3}$ and Thomas Weinacht$^{1}$}
	
	\affiliation{$^{1}$Department of Physics and Astronomy, Stony Brook University, Stony Brook NY 11794-3800}

	\affiliation{$^{2}$Institute of Materials and Environmental Chemistry, Research Centre for Natural Sciences, Hungarian Academy of Sciences, Budapest 1117 Magyar tud\'{o}sok krt. 2, Hungary}

	\affiliation{$^{3}$University of Vienna, Faculty of Chemistry, Institute of Theoretical Chemistry, W\"ahringer Str. 17, 1090 Wien, Austria}
	
	%\thanks{}

\begin{abstract}
We study strong field molecular ionization using few- (four to ten) cycle laser pulses. Employing a supercontinuum light source, we are able to tune the optical laser wavelength (photon energy) over a range of about $\sim$200 nm (500 meV).  We measure the  photoelectron spectrum for a series of different molecules as a function of laser intensity, frequency, and bandwidth and illustrate how the ionization dynamics vary with these parameters. We find that multiphoton resonances and nonadiabatic dynamics (internal conversion) play an important role and result in ionization to different ionic continua.  Interestingly, while nuclear dynamics can be "frozen" for sufficiently short laser pulses, we find that resonances strongly influence the photoelectron spectrum and final cationic state of the molecule regardless of pulse duration -- even for pulses that are less than four cycles in duration.\\
\textit{Preprint of mansucript published in \href
       {http://dx.doi.org/10.1103/PhysRevA.93.051401}{Phys. Rev. A, \textbf{93}, 051401(R) (2016)}.}
\end{abstract}

\maketitle

\section{Introduction}

As ultrafast science progresses to probing attosecond time scales and exploring electronic wavepackets in molecules, it becomes increasingly important to understand the electronic and nuclear dynamics underlying strong field molecular ionization.  High laser intensities of larger than 10$^{12}$ W/cm$^{2}$ can lead to significant dynamic Stark shifting of energy levels, more than the laser bandwidth for $\sim$30 fs or longer pulses \cite{stolow:2005,sussman2006dynamic,trallero2005coherent,trallero2007transition,maeda2011population,SFImodeltheory}.  Strong laser fields can also lead to Freeman resonances \cite{Gibson_MPI,PhysRevLett.59.1092,PhysRevLett.69.1904,freeman:1991} accompanied by nonadiabatic dynamics which couple electronic and nuclear degrees of freedom \cite{DomckePEScalc,mathur2006nonadiabatic,philipp:2011,sandor2016strong}.  For laser pulses with durations greater than 10 fs, ultrafast dynamics such as nuclear motion and internal conversion can take place on excited states of the neutral atom or molecule via multiphoton resonances during ionization, and these can have a strong influence over which states of the cation are accessed \cite{PhysRevLett.86.51,PhysRevLett.109.203007,PhysRevA.86.053406,Tamas:2010,coincidence1}.

A strong field ionization regime which has received considerable attention and offers a simple and intuitive picture of the ionization dynamics is that of quasi-static field or tunnel ionization \cite{seideman1995role,corkum:2011b,walker1994precision,sheehy1998single,guo1999charge}.  For sufficiently high intensities, the ionization rate can be larger than the laser frequency, resulting in ionization which proceeds in an adiabatic fashion such that the instantaneous ionization rate is determined by the instantaneous field. In this limit, the total ionization yield can be calculated by integrating the instantaneous rate for an equivalent static field over the duration of the laser pulse.  This regime is defined by the so called Keldysh parameter having a value much less than one: $\gamma=\omega_{laser}/\omega_{tunnel}<1$ \cite{keldysh1965ionization}. However, this regime is difficult to access practically for multi-cycle pulses with molecules that have relatively low ionization potentials ($\sim 10$ eV). This is because, while the Keldysh parameter may be less than 1 at the peak intensity of the pulse, the ionization yield can be saturated on the rising edge of the pulse before the tunnel regime is reached \cite{lambropoulos1985mechanisms,christov1996nonadiabatic}, and thus most of the ionization yield takes place via multiphoton ionization ($\gamma>1$), in which resonances can play an important role.

The importance of Stark shifted resonances (also known as Freeman resonances) in multiphoton ionization of atoms and diatomic molecules with multicycle pulses ($\sim 30$ fs or longer) has been recognized for quite some time \cite{PhysRevLett.69.1904,mevel1993atoms,fabre1982multiphoton}. For an understanding of this resonance-enhanced ionization, different uses of the term ``adiabatic'' play a role. An electron can move adiabatically both with respect to the laser field, as well as the nuclei of a molecule. In the case of a Freeman resonance, as an intermediate state shifts into resonance, the electron being driven cannot adiabatically follow the laser field, just as a simple harmonic oscillator driven near resonance suffers a phase lag with respect to a driving force. In this sense, the electronic dynamics underlying resonance enhanced ionization are inherently nonadiabatic. This nonadiabatic electronic response can lead to another form of nonadiabatic dynamics. As the intermediate state
will generally not have the same equilibrium geometry as the ground state, the nuclei can begin to move in the excited state. This motion of the nuclei can lead to non-Born-Oppenheimer coupling between different electronic states such that the electron under consideration no longer adiabatically follows the nuclei. In such a case, the electron responds nonadiabatically with respect to both the ﬁeld as well as the nuclei.

These two nonadiabatic effects can together enrich the ionization dynamics and lead to somewhat surprising results. In the limit of very short pulses (i.e., less than four cycles), one expects on the one hand that Freeman resonances no longer play a role because there are insufﬁcient cycles to deﬁne a resonance condition, and on the other hand, that vibrations are ``frozen'' during such a short pulse, minimizing non-Born-Oppenheimer couplings \cite{sandor2016strong}. In this sense, ionization with less than four cycle pulses  should lead to adiabatic ionization dynamcis \cite{brabec:2000,yudin2001nonadiabatic,alnaser2005effects,mathur2008strong}.
Surprisingly, we find that resonances can still play an important role, and thus the ionization is still strongly influenced by nonadiabatic dynamics.

In this Rapid Communication, we measure photoelectrons produced by strong field ionization in a variety of small molecules (with a focus on CH$_2$BrI and CF$_3$I) using both few cycle (9 fs FWHM) optical pulses as well as longer, 30 fs pulses to investigate the importance of dynamic multiphoton resonances and non-adiabatic couplings in the neutral.  In particular, we investigate which states (or what mechanisms) are involved en route to ionization and how they contribute to the final states of the cation.  With longer ($\sim$30 fs) pulses, the central wavelength (photon energy) of the laser light is tuned over a portion of the available optical bandwidth and the photoelectron yield is measured.  We find evidence of ionization to both the ground and excited states of the cation, with resonant enhancement throughout a significant portion of the laser tuning region.

Even for very short (9 fs) pulses, these resonances are still important as long as the laser intensity is sufficiently high to Stark shift the intermediate neutral states into resonance.  This is surprising, since for such a short pulse the resonance condition is only met for a time comparable to an optical cycle, and this is not well defined. From a frequency domain perspective, the laser bandwidth for such a short pulse becomes very broad, and it is not obvious that a few resonance frequencies should play an important role.  We show from solving the time-dependent Schr\"odinger equation for a model system that there is large variation in the multiphoton coupling strength between the ground state and different excited states, and this can contribute to the dominance of one resonance over others.

\section{Experimental Apparatus}

The 9 fs pulses are produced by filamentation in an Argon gas cell with the output of a Ti:sapphire amplifier which produces 1 mJ, 30 fs pulses centered at 785 nm at a 1 kHz repetition rate.  The broad bandwidth produced by the filamentation results in optical radiation spanning an over octave - from about 400 nm to 900 nm at the tails.  The pulse is compressed to near-transfrom limit by using a grating stretcher-compressor system in a 4-f geometry \cite{dugan1997high,fetterman1998ultrafast} and measured using a Self-Diffraction (SD) FROG apparatus \cite{trebino1997measuring}.  After compression, the spectrum produced supports pulses as short as $\sim$6 fs, which are measured with SD FROG to be 8-9 fs.  The 4-f configuration allows the frequency components of the pulse to be manipulated using a variable slit at the focusing element (rather than at the Fourier plane), which avoids hard cuts in the optical spectrum and results in a smooth pulse in the time domain.  Two possible experiments can be conducted with this apparatus: one in which the central wavelength is fixed and the bandwidth is varied, and another where the bandwidth is fixed and the central wavelength is varied.  In the former, we are capable of measuring photoelectron spectra for pulse durations from about 9 fs to about 30 fs.  In the latter, the pulse duration is fixed at about 30 fs and the central wavelength is scanned over $\sim$200 nm from 630 nm (1.95 eV) to 850 nm (1.45 eV) while maintaining a constant photoelectron yield.  The pulse duration was kept roughly constant over the tuning range, but varied slightly from 30 fs due to limited ability to control the slit width and limited pulse energy near the tails of the optical spectrum.

All experiments are performed in a vacuum chamber with a base pressure of 5x$10^{-9}$ torr using an effusive molecular beam at room temperature to produce sample pressures around 5x$10^{-7}$ torr.  The charged particles produced via ionization are accelerated toward a dual stack of microchannel plates (MCPs) and a phosphor screen using an electrostatic lens configured for velocity map imaging (VMI) which produces a two-dimensional projection of the three-dimensional charged particle velocity distribution \cite{Eppink1997RSI}.  The laser is linearly polarized in the plane of the VMI detector.  The photoelectrons and ions may be recorded separately or in coincidence by switching the lens voltages.  Although the experiments described here use only the photoelectrons and are not recorded in coincidence with the ions, previous coincidence measurements are used in assigning the states in the PES \cite{coincidence1}.  The two-dimensional velocity distributions are recorded for each laser shot at 1 kHz using a CMOS camera to capture an image of the particles on the phosphor screen.  A computer algorithm identifies the coordinates of each particle recorded by the camera for each laser shot and synthesizes a background and noise-free single image.  This data is inverse-Abel transformed using the BASEX method \cite{dribinski:2002} and then converted into a photoelectron spectrum.  The following analysis is performed using the photoelectron yield detected at $\pm$30 degrees around the laser polarization direction.  Complete angular integration produces similar features, with reduced contrast of the features of interest.

\section{Results}

\begin{figure}[!]
	\centering
	\includegraphics[width=\columnwidth]{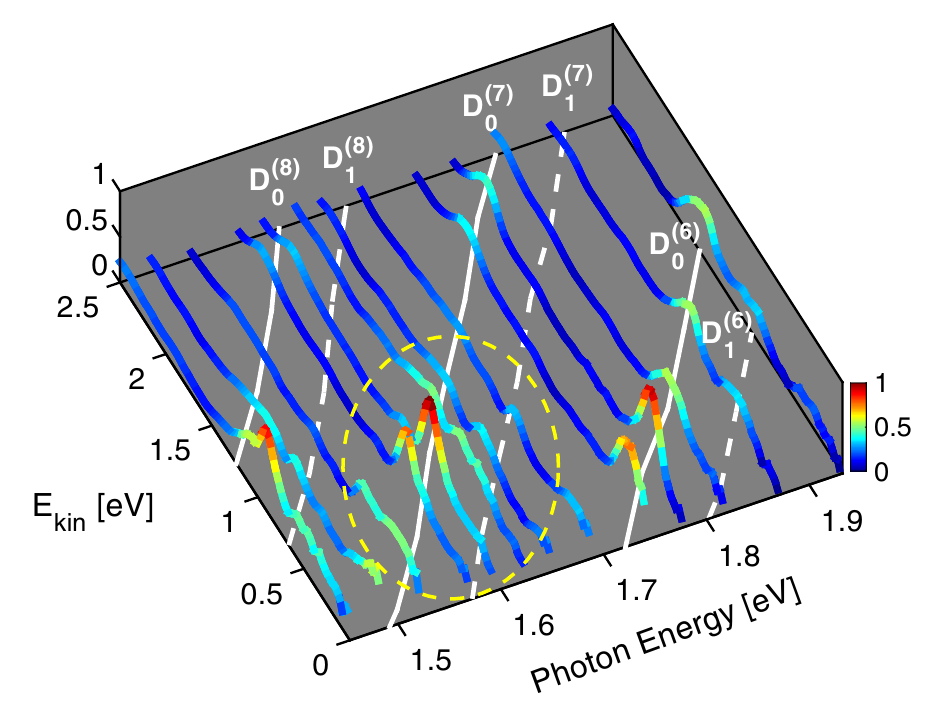}
	\caption{Photoelectron yield (normalized) as a function of photon energy measured in CH$_{2}$BrI.  Each data set is taken with a roughly 30 fs laser pulse whose central wavelength is used to determine the photon energy.  The white solid and dashed lines indicate the expected photoelectron energies for non-resonant ionization of the ionic ground state $D_{0}$ and first excited state $D_{1}$, respectively. These lines are calculated based on the photoelectron yield and laser frequency along with the energy conservation equation described in the text. The superscripts 6, 7, and 8 indicate the number of absorbed photons. Resonance enhanced ionization is highlighted by the dashed yellow circle.}
	\label{cwl_line3D}
\end{figure}

The photoelectron yield as a function of electron kinetic energy and photon energy for ionization of CH$_{2}$BrI is shown in Fig. \ref{cwl_line3D}.  The photoelectron spectra are normalized to the total yield for each photon energy.  The assignment of the peaks in Fig. \ref{cwl_line3D} has been carried out in previous work and verified with electron-ion coincidence measurements \cite{Freeman_control,coincidence1}.  Peaks in the PES which are associated with non-resonant ionization shift as a function of laser frequency and intensity according to $E_{kin}=nh\nu \, - \, I_p\, - \, U_p$.  This equation expresses energy conservation for ionization in a short pulse such that the electrons cannot leave the focal region and regain the ponderomotive energy before the pulse turns off \cite{hertlein1997evidence,bucksbaum1987role}. Here, $I_p$ is the ionization potential ($D_0$=9.69 eV, $D_1$=10.26 eV for CH$_2$BrI \cite{Lago2005JPCA}), $\nu$ is the laser frequency and $U_p$ is the ponderomotive energy, given by: $U_p$=$e^2I/2\epsilon_{0}m_ec\omega^2$, where $I$ is the peak laser intensity and $\omega$ is the angular frequency.  As the total photoelectron yield goes as $I^n$ for $n$ absorbed photons, we use the yield to calculate the ponderomotive shift for each photon energy assuming $\approx$ 0.7 eV ponderomotive shift at 1.45 eV (the lowest photon energy used in Fig. \ref{cwl_line3D}).  The solid and dashed white lines indicate the expected peak positions for non-resonant ionization by absorption of a minimum of $n$-photons along with the calculated ponderomotive shift.  Peaks lying along these lines are due to non-resonant multiphoton ionization by absorption of $n$-photons to the ground ionic state, $D_{0}$, and to the first excited state, $D_{1}$.  Peaks occurring with energies corresponding to $n+m$ photon absorption for $m\geq$ 1 are due to above threshold ionization (ATI).  Across the photon energy range available, most of the ionization takes place to the ground ionic state, $D_0$, by absorbing 6-to-8 photons. Some ionization also leaves the molecule in $D_{1}$.

Fig. \ref{cwl_line3D} reveals resonant and non-resonant features.  The PES when the photon energy is off-resonance can be seen from 1.75-1.95 eV.  Here, the $D_0$ and $D_1$ peaks lie along the predicted positions and shift as expected with laser frequency (wavelength).  Peaks which do not shift linearly with laser frequency and do not lie on top of the white lines on the graph are those which are resonantly enhanced and come from ionization off of the peak of the laser pulse intensity and thus have a lower $U_p$, leading to a different peak position than one would calculate for nonresonant ionization.   Deviations from the predicted energies can be observed at $E_{kin}$ $\approx$ 0.7 eV for photon energies of 1.55 - 1.68 eV around the 7$^{th}$ photon order and are highlighted in the dashed yellow circle.  In particular, the peak at $E_{kin}$ $\approx$ 0.7 eV at a photon energy 1.63 eV is lower than the $E_{kin}$ expected for non-resonant ionization to $D_{0}$, but higher than expected for non-resonant ionization to $D_{1}$.   This peak comes from resonantly enhanced ionization - an intermediate state is Stark shifted into resonance during the pulse (i.e. a Freeman resonance), enhancing the ionization yield and leading to ionization coming principally at an intensity where the intermediate state is resonant \cite{Freeman_control}.  Thus, we can identify resonant features in Fig. \ref{cwl_line3D} as peaks lying at higher energies than those predicted by the energy conservation equation for a full ponderomotive shift (at the peak intensity).

\begin{figure}[!]
	\centering
	\includegraphics[width=\columnwidth]{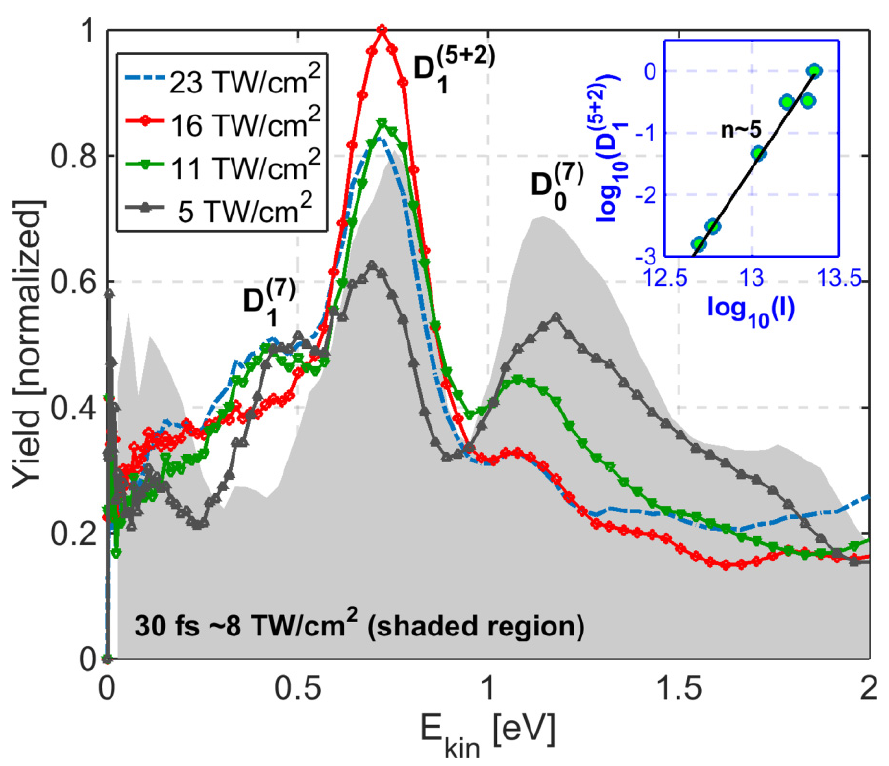}
	\caption{Photoelectron spectra as a function of laser intensity in CH$_{2}$BrI produced from ionization with a 9 fs broadband pulse.  The data are individually normalized to the total number of detected electrons from 0-2 eV so the peak positions may be compared.  The highest intensity used is shown as a dashed line in the PES (23 TW/cm$^{2}$).  The shaded data is the PES produced by a 30 fs pulse with photon energy of 1.65 eV as shown in Fig. \ref{cwl_line3D} for comparison.  Inset:  Logarithm of the resonantly enhanced $D_1$ yield vs logarithm of intensity with a slope of $\approx$5.}
	\label{bimint}
\end{figure}

In order to characterize the resonance around 1.6 eV, the PES resulting from ionization of CH$_{2}$BrI for a series of different intensities are shown in Fig. \ref{bimint}.  The spectra shown with colored lines are for ionization with 9 fs pulses, while the shaded grey spectrum is for a 30 fs pulse for comparison.  All PES are normalized to the integrated yield from 0-2 eV.  The peak at $\sim 0.7$ eV (labeled $D_1^{(5+2)}$) does not shift with intensity, consistent with resonantly enhanced ionization at intensities above 7 TW/cm$^2$.  The 5+2 label indicates that it is a 5 photon resonance, based on the intensity dependent yield shown in the inset.  The peak at $\sim$1.2 eV slowly disappears as a function of increasing intensity for 9 fs pulses, although this does not happen for 30 fs pulses \cite{sandor2016strong}.  As discussed in a previous work \cite{sandor2016strong}, this is due to a subtle interplay between resonant enhancement and internal conversion between intermediate states - both nonadiabatic effects during the ionization dynamics.  Although the internal conversion is suppressed for 9 fs pulses, which are shorter than the shortest vibrational period in CH$_{2}$BrI, resonant enhancement surprisingly persists, despite the fact that 9 fs pulses correspond to less than four optical cycles.

While the ionization for intensities above $\sim$7 TW/cm$^2$ is dominated by resonant enhancement, there are non-resonant contributions to $D_{1}$ around $E_{kin}$ = 0.4-0.5 eV and to $D_0$ around 1.2 eV.  At intensities below $\sim$7 TW/cm$^2$,  the peaks shift ponderomotively to lower kinetic energies as the intensity increases until about 7 TW/cm$^{2}$, where $D_1$ becomes resonantly enhanced.  The $D_0$ and $D_1$ peaks stop shifting near the resonance intensity.  The non-resonant contributions from $D_{0}$ and $D_{1}$ become reduced as the intensity is increased and the $D_{1}$ resonance becomes the dominant contribution to the PES.

\begin{figure}[!]
	\centering
	\includegraphics[width=\columnwidth]{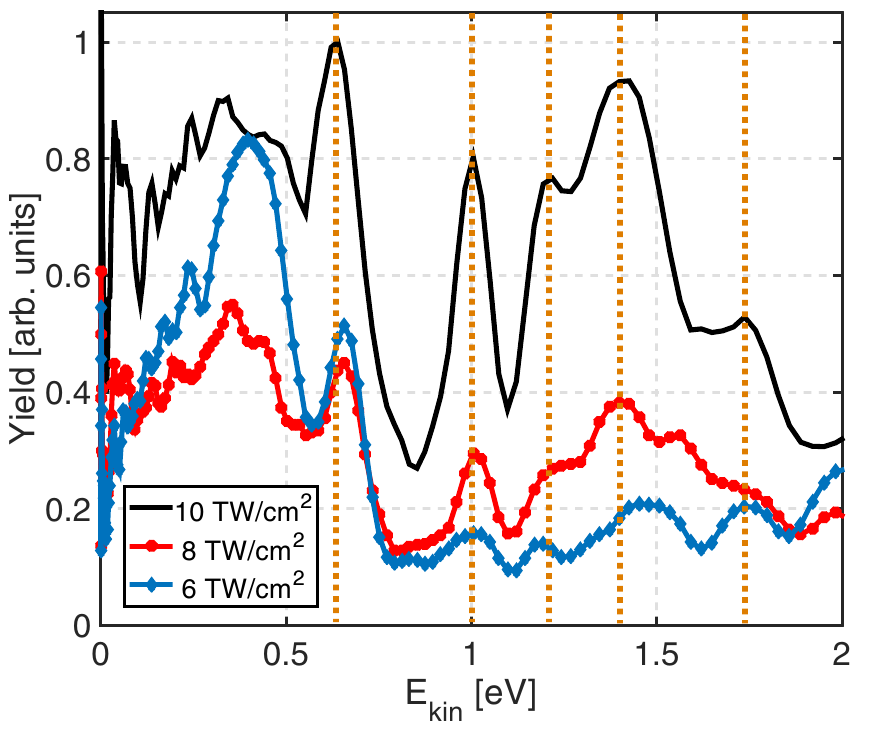}
	\caption{Photoelectron spectrum for CF$_{3}$I$^{+}$ as a function of intensity in TW/cm$^{2}$ produced with a 9 fs pulse from 6-10 TW/cm$^2$.  The dotted vertical lines mark positions of resonant peaks.}
	\label{cfpeaks}
\end{figure}

Similar measurements on a range of molecules (including CH$_2$BrCl, C$_4$H$_6$, C$_6$H$_8$, C$_{10}$H$_{16}$, C$_6$H$_5$I, CS$_2$, and CF$_3$I) show evidence for resonantly enhanced ionization playing an important role in the ionization dynamics and photoelectron yield: all of these molecules have peaks which dominate the PES and do not shift with laser intensity.  An example of the resonant features observed in CF$_3$I is shown in Fig. \ref{cfpeaks}.  The figure shows the PES for different intensities using 9 fs pulses.  Multiple resonant peaks, whose positions do not shift with intensity can be seen.  This clearly demonstrates that even for broadband 9 fs pulses, which contain less than four cycles, resonance enhancement can still dominate the ionization yield.

While initially surprising, this can be understood in terms of a simple time domain picture of the laser molecule interaction. In the simplest case of two states (g and e) multiphoton coupled by an unshaped (transform limited) strong laser field, the equations for the state amplitudes, $b_g$ and $b_e$, can be written as \cite{SFImodeltheory}:
\begin{align}
i\dot{b}_g &= \Omega_{eg}(t)e^{i\alpha(t)}b_e \nonumber\\
i\dot{b}_e &= \Omega_{eg}(t)e^{-i\alpha(t)}b_g \nonumber\\
\label{eqn:final3}
\end{align}
where $\Omega_{eg}$ is the multiphoton Rabi frequency (including both electronic and vibrational factors), and the molecule-field phase, $\alpha(t)$, is given by:
\begin{align}
\alpha(t) = \Delta_{eg}t - \int_{-\infty}^t\delta_\omega^{(s)}(t')dt' \nonumber\\
\end{align}
Here $\delta_\omega^{(s)} \equiv \omega^{(s)}_e - \omega^{(s)}_g \propto I(t)$ is the dynamic Stark shift between ground and excited states and $\Delta_{eg}$ is the field free multiphoton detuning. In this picture, the resonance condition can be understood in terms of a slowly varying $\alpha(t)$ kept near zero for many cycles of the driving pulse.  However, for a short pulse, population cannot build up in the excited state over many cycles. In order for the excited state to resonantly enhance the ionization, the multiphoton Rabi frequency, $\Omega_{eg}$, must therefore be sufficiently high such that significant population is transferred to/through the excited state in an optical cycle.

\begin{figure}[!]
	\centering
	\includegraphics[width=\columnwidth]{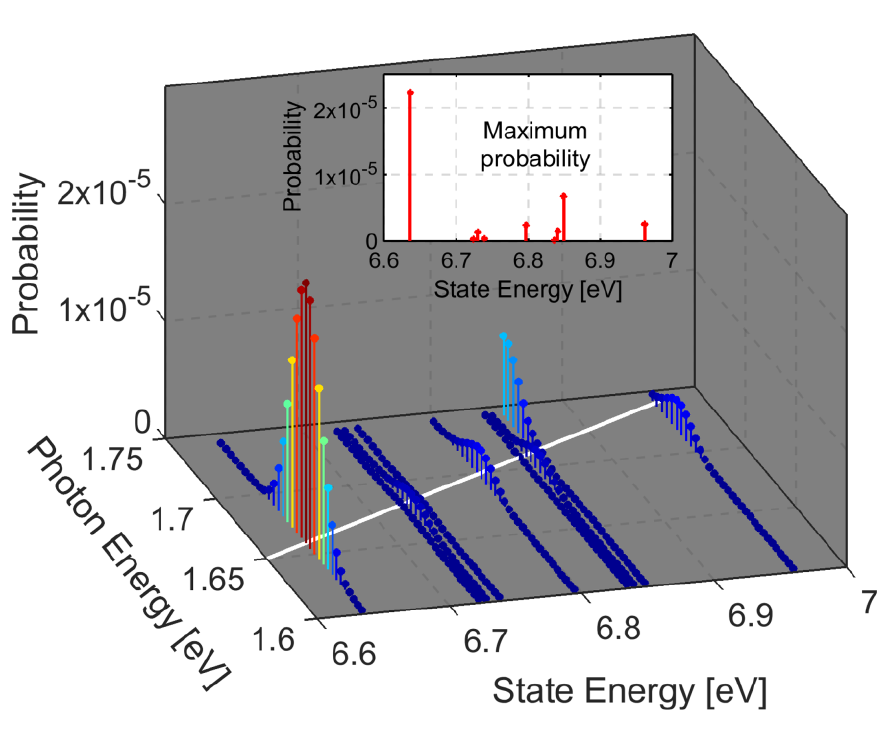}
	\caption{Transition probabilities from solving the time-dependent Schr\"odinger equation for a model system (see text for discussion) with 68 neutral excited states with a 30 fs laser pulse duration and an intensity of 10 TW/cm$^2$. The state energy is the energy of each neutral state relative to the ground state. The photon energy is tuned over a small range such that multiple states come into resonance. The solid white line indicates the energies of a four-photon resonance \cite{Note1}. The inset shows the maximum probability amplitude when states are near four-photon resonance.}
	\label{sim}
\end{figure}

A natural question which arises is what determines the  coupling strength of these states to the ground state: vibrational wave function overlap (Franck Condon factors), or multiphoton coupling strength?  Calculations of population transfer for displaced potentials (with displacements taken from electronic structure calculations for excited states of CH$_2$BrI) suggest that Franck Condon factors are not what determine which electronic states play an important role, since the population transfer to excited states varies relatively slowly with displacement between the ground and excited state potentials for typical displacements.

In order to see the extent multiphoton Rabi frequency variation, we solved the time dependent Schr\"odinger equation for a model system based on CH$_2$BrI, such that the energy spectrum and coupling strengths are realistic, although the limited number of states used in the calculation means that the model cannot be considered to truly represent CH$_2$BrI itself. This is more straightforward than trying to calculate the multiphoton coupling strengths explicitly, which would involve very large sums over off resonant states of the molecule \cite{trallero2005coherent}.  Fig. \ref{sim} shows the results (final state populations) of solving the time-dependent Schr\"odinger equation with fixed nuclei using 68 excited states for a 30 fs laser pulse with peak intensity of 10 TW/cm$^2$.  The state energies and transition dipole matrix used in the simulation were calculated at the level of multi-state complete-active-space perturbation theory of second order \cite{finley1998multi} at the Franck-Condon geometry with spin-orbit coupling included \cite{Note2}.  The shading is proportional to the final population in each state at the end of the pulse, which provides a measure of the multiphoton coupling strength.  The probability is calculated as a function of laser photon energy from 1.55 to 1.75 eV in the spirit of Fig. \ref{cwl_line3D}.  The solid white line marks a four-photon resonance based on the photon energies.

We find that there is a large probability (i.e., population transferred to states) near the four-photon resonance as the states around the 6-7 eV level Stark shift into resonance \cite{Note1}.  The inset of Fig. \ref{sim} shows the population transferred to each state when on four-photon resonance as the laser frequency is varied.  Interestingly, the multiphoton coupling strength is much stronger for certain states, even though each state shifts through resonance.  The large variation in these couplings is an indication of how specific excited states may be more strongly coupled to the ground state than others via the strong field of the laser pulse.  Based on these calculations, we conclude that multiphoton coupling strengths vary widely and are difficult to predict \textit{a priori}.

\section{Conclusion}

In conclusion, we find that resonances play an important role in strong field molecular ionization, even in the limit of pulses less than four cycles in duration. These resonances can be mapped out and identified by studying the ionization yield as a function of laser frequency.  Calculations which solve the TDSE for a large number of states indicate that multiphoton coupling strengths can vary greatly for states with similar energy allowing ionization to excited states of the cation to dominate the ionization yield.

\begin{acknowledgments}
This work has been supported by the National Science Foundation under award number 1205397 and the Austrian Science Fund (FWF) through project P25827. Support from the European XLIC COST Action 1204 is also acknowledged.
\end{acknowledgments}

%\bibliographystyle{apsrev4-1}
%\bibliography{shortbib4_VT}


\end{document}